\documentclass[11pt,twoside]{article}

\usepackage{asp2006}
\usepackage{epsf}
\usepackage{psfig}
\usepackage{lscape}

\begin{document}
\title{X-Ray Nuclei in Radio Galaxies: Exploring the Roles of Hot and Cold Gas Accretion}   %%% Fill in title
\author{D. A. Evans\altaffilmark{1}, M. J. Hardcastle\altaffilmark{2}, J. H. Croston\altaffilmark{2}}   %%% Fill in author names
\altaffiltext{1}{Harvard-Smithsonian Center for Astrophysics, 60 Garden Street, Cambridge, MA 02138}    %%% Fill in author affiliations
\altaffiltext{2}{School of Physics, Astronomy \& Mathematics, University of Hertfordshire, College Lane, Hatfield, AL10 9AB, UK}

\begin{abstract} %%% Abstract to run on from here.

We present results from {\it Chandra} and {\it XMM-Newton} spectroscopic observations of the nuclei of $z<0.5$ radio galaxies and quasars from the 3CRR catalog, and examine in detail the dichotomy in the properties of low- and high-excitation radio galaxies. The X-ray spectra of low-excitation sources (those with weak or absent optical emission lines) are dominated by unabsorbed emission from a parsec-scale jet, with no contribution from accretion-related emission. These sources show no evidence for an obscuring torus, and are likely to accrete in a radiatively inefficient manner. High-excitation sources (those with prominent optical emission lines), on the other hand, show a significant contribution from a radiatively efficient accretion disk, which is heavily absorbed in the X-ray when they are oriented close to edge-on with respect to the observer. However, the low-excitation/high-excitation division does not correspond to the FRI/FRII division: thus the Fanaroff-Riley dichotomy remains a consequence of the interaction between the jet and the hot-gas environment through which it propagates. Finally, we suggest that accretion of the hot phase of the IGM is sufficient to power {\it all} low-excitation radio sources, while high-excitation sources require an additional contribution from cold gas that in turn forms the cold disk and torus. This model explains a number of properties of the radio-loud active galaxy population, and has important implications for AGN feedback mechanisms.

\end{abstract}

\section{Unified models and radio-loud AGN: the excitation dichotomy}

In standard AGN models, efficient disk accretion of cold matter on to the central supermassive black hole provides the radiation field that photoionizes the optical broad-line region (BLR) and narrow-line region (NLR) and gives rise to X-ray emission via Compton scattering. Without radiatively efficient accretion via the disk, none of these standard features of such an AGN would be observed. Unified models propose that a direct view of the BLR and the optical continuum may be obscured (e.g., in Seyfert 2s) by a dusty `torus': but in this case the torus re-radiates strongly in the mid-IR band, so that the presence of a luminous AGN can still be inferred.

By analogy with radio-quiet objects, we would expect that face-on radio-loud objects (the broad-line radio galaxies and radio-loud quasars) would show both broad and narrow optical lines, while edge-on radio-loud objects (narrow-line radio galaxies, NLRG) would show only narrow optical lines, and would have a clear mid-infrared signature of the absorbing torus. This radio-loud unified scheme well describes the nuclei of many of the most powerful radio sources (e.g., \citealt{bar89,haas04}), although it is well known that many radio galaxies do not have the strong optical line-emission that is expected from a conventional AGN (\citealt{hine79,jr97}). The objects lacking these narrow lines, the low-excitation radio galaxies (LERGs), in general show no evidence in the mid-IR for an obscuring torus, either at low or high luminosities (\citealt{why04,ogle06}). Most low radio-power (FRI) sources galaxies are LERGs, whereas most high radio-power (FRII) sources are high-excitation radio galaxies (HERGs -- i.e., NLRGs, BLRGs, and quasars). However, {\bf the low-excitation/high-excitation division does not correspond to the FRI/FRII division}: there is a small number of high-excitation FRI sources (including the nearest radio-loud AGN, Centaurus A), as well as a significant population of low-excitation, radio-powerful FRII sources.

\section{The origin of X-ray emission}

\subsection{Overview}
The physical origin of nuclear X-ray emission in radio-loud AGN has been a topic of considerable debate. In particular, it has been unclear as to whether the emission primarily originates in an quasi-isotropic accretion flow, or instead is associated with an intrinsically beamed parsec-scale radio jet. The detection with {\it ROSAT} of unabsorbed, power-law X-ray emission in the B2 (\citealt{can99}) and 3CRR (\citealt{har99}) samples, together with observed correlations between both the X-ray and VLA radio core fluxes and luminosities led those authors to suggest a nuclear jet-related origin for at least the soft X-ray emission. However, the torus required to obscure the optical and UV disk emission should also obscure any X-ray emission associated with the accretion disk. Therefore, AGN inclined at low to intermediate angles with respect to the observer should show a component of heavily absorbed, accretion-related, nuclear X-ray emission, similar to that observed in Seyfert 2 galaxies. This was borne out by early studies of individual objects (e.g., \citealt{uen94}) as well as more detailed studies of large samples with hard X-ray instruments such as {\it ASCA} and {\it BeppoSAX} (e.g., \citealt{sam99,gra06}).

{\it Chandra} and {\it XMM-Newton} have revolutionized the study of radio-galaxy nuclei. {\it Chandra} is particularly suited to this task, owing to its high angular resolution and corresponding ability to spatially separate AGN emission from that of the surrounding hot-gas environment. In turn, the excellent sensitivity of {\it XMM-Newton} allows us to search for, with greater sensitivity, components of heavily absorbed (and likely accretion-related) X-ray emission. As the orientation-dependent effects of relativistic beaming and the putative obscuring torus are expected to play a large part in determining the observed properties of a radio-galaxy nucleus, it is important to select sources based on their low-frequency (and hence isotropic) emission characteristics, such as in the 3C and 3CRR samples. In these proceedings, we concentrate on {\it Chandra} and {\it XMM-Newton} observations of radio galaxies drawn from the 3CRR catalog in the redshift range $z<0.5$, as discussed by \cite{don04}, \cite{evans06}, \cite{bal06}, and \cite{hec06}. Of the 86 $z<0.5$ 3CRR radio sources, 40 have been observed with {\it Chandra} or {\it XMM-Newton}.

\subsection{X-ray spectra and correlations}

Spectral analysis with {\it Chandra} and {\it XMM-Newton} of the $z<0.5$ sources shows that {\it every one} possesses an unabsorbed component of nuclear X-ray emission, as first seen with {\it ROSAT}. However, narrow-line (i.e., high-excitation) radio galaxies (almost all of which have FRII morphologies, with a handful showing FRI radio features) show an {\it additional} heavily absorbed nuclear component, with a column density often in excess of $10^{23}$~cm$^{-2}$, accompanied by narrow Fe K$\alpha$ line emission. The properties of these narrow-line FRIIs are thus consistent with the expectation from unified models. However, {\it no} low-excitation radio galaxy in the 3CRR sample shows any evidence for this type of heavily absorbed nuclear component (\citealt{hec06}).

\begin{figure}[t]
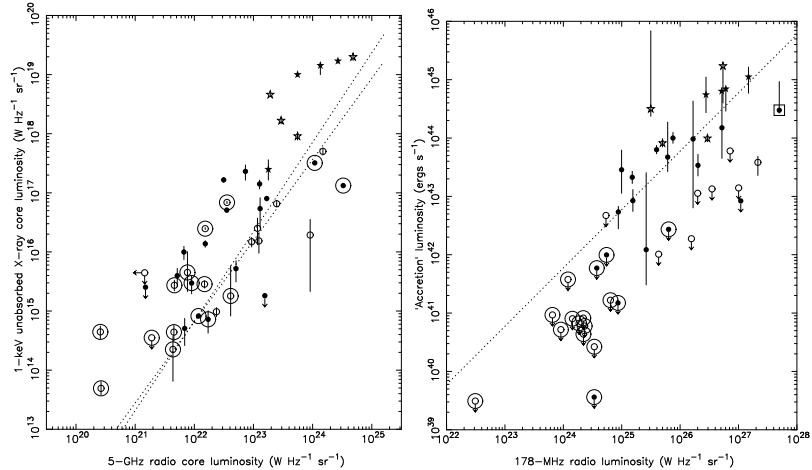

\begin{center}
\epsfxsize 5.2cm
\epsfbox{devans_radio-xray.ps}
\epsfxsize 5.3cm
\epsfbox{devans_lum-lum.ps}
\caption{{\it (a)} X-ray luminosity of the unabsorbed X-ray component as a function of 5-GHz radio core luminosity. Open circles are LERG, filled circles NLRG, open stars BLRG, and filled stars quasars. Surrounding circles mean a source is an FRI. Dotted lines show 90\% confidence regression lines fitted through the NLRGs and LERGs. BLRGs and quasars lie above the line. {\it (b)} X-ray luminosity of the accretion-related component for the combined $z<0.5$ sample as a function of 178-MHz total radio luminosity. The regression line is determined by the NLRG.}
\label{radio-xray}
\end{center}
\end{figure}

Figure~\ref{radio-xray}a shows the 1-keV luminosity of the {\it unabsorbed} power-law component against the 5-GHz luminosity of the radio core. This illustrates the correlation between the soft X-ray emission and the radio core, as first discussed by \cite{har99}. The correlation implies that the X-ray emission is affected by relativistic beaming in the same manner as the radio, and so suggests a physical relationship between the two, most plausibly at the base of the jet. There is no systematic difference in the behavior of FRIs and FRIIs, or NLRGs and LERGs. The broad-line objects in the sample lie above the trendline established by the NLRG and LERG, and \cite{evans06} attribute this to the presence of unabsorbed accretion-related emission in their spectra, consistent with the near face-on orientation of these objects in unified AGN models.

\section{The ubiquity of the torus and constraints on the accretion mode}

We have established that low-excitation radio galaxies possess no evidence for heavily absorbed nuclear X-ray emission. What does this imply for the ubiquity of the torus and the nature of the accretion flow? We assumed that in each LERGs, there exists a `hidden' component of accretion-related emission that is obscured by a column $10^{23}$ cm$^{-2}$, in addition to jet-related component of X-ray emission that dominates the spectrum. We then determined the 90\%-confidence upper limit to the 2--10 keV luminosity of this hidden component. The accretion-related luminosity of the HERGs is given by the unabsorbed luminosity of the heavily obscured emission in NLRGs, and by the offset (Fig.~\ref{radio-xray}a) in the case of broad-line objects.

Figure~\ref{radio-xray}b shows a plot of the unabsorbed 2--10 keV accretion-related luminosity against 178-MHz radio luminosity, for both the low- and high-excitation radio sources.  The upper limits on the accretion-related components in the LERGs, given our assumed absorbing column of $10^{23}$ cm$^{-2}$, lie systematically below the detected HERGs at all radio powers. If no obscuring region is present at all in LERGs, then the luminosity of any accretion-related emission will be substantially lower than that shown.

We now turn to the implications of this result on the nature of the accretion flow in low- and high-excitation sources. One widely discussed model (e.g.,~\citealt{rey96,don04}) is that there exists a fundamentally different accretion {\it mode} in FRI- and FRII-type sources, such that the accretion-flow luminosities and radiative efficiencies of FRI-type radio galaxies are systematically lower than those of FRII-type radio galaxies. The main difficulty from the point of view of radio-galaxy physics with such a model is that the FRI/FRII dichotomy, and its dependence on host galaxy properties (\citealt{led96}), can be explained purely in terms of jet power and the interaction with the environment (e.g.,~\citealt{bic95}). However, we can readily modify the accretion model to accommodate a scheme in which {\bf radiatively inefficient accretion flows power LERGs, and efficient accretion via a standard thin disk powers HERGs.}

\section{Powering AGN: Hot vs. cold gas accretion}

\subsection{Bondi accretion}

We have argued that LERGs may be a class of luminous active galaxies that accrete radiatively inefficiently, with almost all the available energy from accretion being channeled into the jets, while high-excitation sources are powered by radiatively efficient accretion. In \cite{hec07} we considered whether these different accretion modes may be a result of a different {\it source} for the accreting gas, building on the recent result of \cite{allen06}, who showed that some low-luminosity radio galaxies in the centers of clusters could be powered by Bondi accretion from the hot, X-ray emitting medium.

We can use X-ray observations of the hot-gas environment of radio sources to estimate the Bondi accretion rate, and in turn constrain the amount of power that the central supermassive black hole can extract from accretion of the hot phase of the IGM.  The Bondi rate is given by $\dot M = \pi\rho_{\rm A} G^2 M_{\rm BH}^2 / c_s^3$, where $r_A$ is the Bondi accretion radius and $c_{\rm s}$ is the sound speed in the medium. The available power for AGN activity, $P_{\rm B} = \eta \dot M c^2$, where $\eta$ is an efficiency factor (assumed to be 0.1). To derive black-hole masses for radio galaxies we use the relationship between $M_{\rm BH}$ and K-band absolute bulge magnitude derived by \cite{mar03} for nearby sources. To show that the jet can be powered by accretion of the IGM, we require that the (kinetic plus radiative) jet power $Q\la P_{\rm B}$. In \cite{hec07}, we presented a detailed description of methods used to estimate the jet power. In short, we used the \cite{wil99} relation between jet power and 151-MHz luminosity density: $Q_W=3\times10^{38} f^{3/2} L^{6/7}_{151}\ {\rm W}$, where $f$ parametrizes our ignorance of true jet powers.

\begin{figure}
\begin{center}
\epsfxsize 8cm
\epsfbox{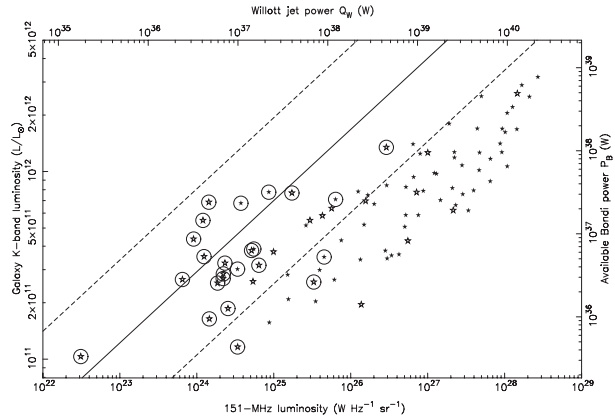}
\caption{K-band host galaxy luminosity against 151-MHz luminosity for LERGs and HERGs, with conversions into Willott jet power ($Q_{\rm W}$) and available Bondi power ($P_{\rm B}$). Open stars are LERGs and filled stars are NLRGs. A circle round a data point indicates an FRI.  The central solid line shows equality between the predicted Bondi power and the Willott jet power, and the dashed lines are separated from the solid line by one order of magnitude.}
\label{bondi-plot}
\end{center}
\end{figure}

In Figure~\ref{bondi-plot} we plot the observational quantities, radio luminosity and $K$-band luminosity, together with their conversion to Bondi power and jet power. Figure~\ref{bondi-plot} shows that the nearby FRI radio galaxies almost all lie within a factor of a few of the line of $Q_W = P_{\rm B}$. Secondly, it shows that the majority of low-excitation FRII radio galaxies in our sample also lie close to this line. And thirdly, it shows that there is a population of FRII sources, encompassing most of the narrow-line FRII sources (and therefore, presumably, all high-excitation sources), that have jet powers exceeding the available Bondi powers (for our choice of central gas properties), often by more than two orders of magnitude.

In summary, we have shown that it is possible that all the low-excitation radio galaxies are powered by accretion from the hot phase, consistent with the fact that their nuclear spectra show no evidence for cold material close to the nucleus (i.e., no evidence for the `torus'). On the other hand, we have seen that narrow-line radio galaxies (and therefore other high-excitation sources) have clear evidence for accretion disks and tori, cannot be powered in this way --- the large amounts of cold material in the nucleus and the radiative efficiency of accretion are naturally explained if these objects are powered by accretion of cold material via a thin disk in the standard manner.

\subsection{Implications}

\cite{hec07} present a detailed discussion of the implications of hot vs. cold gas accretion for radio sources. We summarize these below:

\smallskip\noindent
\large {\bf Feedback:} \normalsize AGN feedback is thought to be an important aspect of galaxy formation models (e.g.,~\citealt{cro06,bow06}). An important feature of these models is that the AGN should both be able to influence, {\it and should be influenced by}, the X-ray emitting phase. Direct accretion of the hot phase provides an natural way of ensuring that the AGN activity is regulated by the gas properties at the cluster center. However, this is only possible for a `hot-mode' radio source. Cold-mode sources do not have this direct connection between the hot phase and the rate of fueling of the AGN: instead, the jet power is controlled solely by the accretion rate of cold gas, and so these sources can potentially inject a significant amount of energy input to the IGM without regulation from their hot-gas environments.

\smallskip\noindent
\large {\bf Environments:} \normalsize In the model we have outlined we expect different types of active galaxies to be found in different environments. Cold-mode accretion requires a supply of cold gas: the easiest way for an elliptical galaxy to acquire this is by a merger with a gas-rich system. Samples of high-excitation radio galaxies should thus show evidence for mergers and interactions, as is often observed (e.g.,~\citealt{hec86}). Host galaxies of cold-mode systems do not need a rich environment, or to be at the bottom of a deep potential well, so long as galaxy-galaxy mergers can take place. By contrast, hot-mode accretion requires a supply of hot gas and a massive central black hole. Both the black hole mass and the mass of the galaxy-scale X-ray halo (e.g.,~\citealt{mat03}) are correlated with the mass of the host galaxy. Thus we expect hot-mode systems -- which, observationally, include almost all FRI radio galaxies -- to favor massive galaxies, and the most powerful radio sources to tend to be group- or cluster-dominant systems (e.g., \citealt{lon79,best04}).

\noindent An overview of the properties of LERGs and HERGs is given in Table~\ref{summary}

\begin{table}[t]
\begin{tabular}{p{2.5cm}p{5cm}p{5cm}}
\hline
 & Low-excitation (LERG) & High-excitation (HERG) \\
\hline
Definition & No narrow optical line emission. & Prominent optical emission lines, either narrow (NLRG) or broad (BLRG), or quasar. \\ \hline

Fanaroff-Riley classification & Almost all FRIs are LERGs, as well as a significant population of FRIIs. & Most FRIIs are HERGs, as are a handful of FRIs (e.g., Cen~A). \\ \hline

X-ray spectra & Jet-related unabsorbed power law only. Upper limits only to 'hidden' accretion-related emission. & Jet-related unabsorbed power law + significant accretion contribution (heavily absorbed in NLRGs). \\ \hline

Accretion-flow type & Highly sub-Eddington. Likely radiatively inefficient. & Reasonable fraction of Eddington. Likely standard accretion disk. \\ \hline

Optical constraints & Strong radio/optical/soft X-ray correlations. Optical emission is jet-related. & Strong radio/optical/soft X-ray correlations. Optical emission is jet-related. \\ \hline

Fueling mechanism & Bondi accretion of hot ISM. & Additional fuel supply needed, likely cold gas. \\ \hline

Implications of AGN fueling & Significant feedback between AGN and environment. & Potentially large energy input to IGM, decoupled from hot-gas environment. \\ \hline

\hline
\end{tabular}
\label{summary}
\caption{Overview of the properties of low- and high-excitation radio galaxies}
\end{table}

\acknowledgements D.A.E. gratefully acknowledges support from NASA through {\it XMM-Newton} award NNX06AG37G.

\end{document}